\documentclass[
    12pt            
  ]
  {article}
\usepackage{a4}
\usepackage{amsfonts}
\usepackage{amsmath}
\usepackage{amssymb}
\usepackage{lscape}
\usepackage{epsfig}
\usepackage{multirow}

\newcommand{\drawsquare}[2]{\hbox{%
\rule{#2pt}{#1pt}\hskip-#2pt
\rule{#1pt}{#2pt}\hskip-#1pt
\rule[#1pt]{#1pt}{#2pt}}\rule[#1pt]{#2pt}{#2pt}\hskip-#2pt
\rule{#2pt}{#1pt}}
\newcommand{\Ysymm}{\raisebox{-.5pt}{\drawsquare{6.5}{0.4}}\hskip-0.4pt%
         \raisebox{-.5pt}{\drawsquare{6.5}{0.4}}}
\newcommand{\Yasymm}{\raisebox{-3.5pt}{\drawsquare{6.5}{0.4}}\hskip-6.9pt%
        \raisebox{3pt}{\drawsquare{6.5}{0.4}}}

\newcommand{\be}{\begin{equation}}
\newcommand{\ee}{\end{equation}}
\newcommand{\ba}{\begin{array}}
\newcommand{\ea}{\end{array}}
\newcommand{\bea}{\begin{eqnarray}}
\newcommand{\eea}{\end{eqnarray}}

\newcommand{\ov}{\overline}
\def\IR{\relax{\rm I\kern-.18em R}}

\def\IP{\relax{\rm I\kern-.18em P}}
\def\inbar{\vrule height1.5ex width.4pt depth0pt}
\def\IC{\relax\,\hbox{$\inbar\kern-.3em{\rm C}$}}

\def\K3{{\bf K3}}

\def\ov{\overline}

\def\n2d{\cN_{V^*}^{\otimes 2}}

\def\IC{\mathbb{C}}

\def\IR{\mathbb{R}}

\def\IP{\mathbb{P}}

\def\cN{{\mathcal N}}

\begin{document}






\title{
\begin{flushright} \vspace{-2cm}
{\small UPR-1212-T\\
ROM2F/2009/20\\
} \end{flushright} \vspace{1.5cm}
Mass Hierarchies vs. Proton Decay in MSSM Orientifold Compactifications\footnote{The work is based in part  on talks given at DPF `09 (M.C.), String Phenomenology `09 (M.C., R.R.), Supersymmetry `09 (M.C.)  and Galileo Galilei Institute School on String Theory (M.C.).}}
 \vspace{1.0cm}
\author{\small  Mirjam Cveti{\v c}$^1$, James Halverson$^1$,  Robert Richter$^2$}

\date{}

\maketitle
\begin{center}
\emph{${}^1$ Department of Physics and Astronomy, University of Pennsylvania, \\
     Philadelphia, PA 19104-6396, USA }\\
\emph{${}^2$ Dipartimento di Fisica and Sezione I.N.F.N. \\ Universit\`a di Roma ``Tor Vergata''\\
Via della Ricerca Scientifica, 00133 Roma, Italy}\\
\vspace{0.5cm}
\tt{cvetic@cvetic.hep.upenn.edu, jhal@sas.upenn.edu, rrichter@roma2.infn.it}
\vspace{0.2cm}
\end{center}
\vspace{0.2cm}

\maketitle

\begin{abstract}
We review two systematic bottom-up analyses of MSSM quivers recently performed by the authors.  We extend the analysis of \cite{Cvetic:2009yh} by including constraints arising from proton decay via dimension $5$ operators and present all four-stack quivers in the Madrid embedding which satisfy this additional constraint. Furthermore, we investigate and make precise the interplay between mass hierarchies obtained via factorizable Yukawa textures and the presence of dimension $5$ proton decay operators in MSSM orientifold compactifications. We discuss this issue in a five-stack quiver, first presented in \cite{Cvetic:2009new}, which exhibits proper mass hierarchies and no rapid proton decay.
\end{abstract}

\thispagestyle{empty} \clearpage

\section{Introduction} \label{sec introduction}

There have been extensive efforts \cite{Ibanez:2008my,Leontaris:2009ci,Anastasopoulos:2009mr,Cvetic:2009yh,Cvetic:2009new,Kiritsis:2009sf}, recently, to construct semi-realistic bottom-up MSSM quivers realized by intersecting D-branes (and their T-dual pictures)\footnote{For reviews on this subject, see \cite{Blumenhagen:2005mu, Marchesano:2007de,Blumenhagen:2006ci}.}. In these compactifications, the gauge groups arise from stacks
of D6-branes that fill out four-dimensional spacetime and wrap three-cycles in the internal Calabi-Yau threefold. Chiral matter arises at the intersection of two different D6-brane stacks in the internal space, and the multiplicity of the chiral matter is given by the topological intersection number of the respective three-cycles.

Once the MSSM spectrum has been realized\footnote{For
original work on globally consistent non-supersymmetric intersecting D-branes, see
\cite{Blumenhagen:2000wh,Aldazabal:2000dg,Aldazabal:2000cn,Blumenhagen:2001te},
and for chiral globally consistent  supersymmetric ones, see
\cite{Cvetic:2001tj,Cvetic:2001nr}. For supersymmetric MSSM realizations, see \cite{Honecker:2004kb,Gmeiner:2007zz,Gmeiner:2008xq},
and for supersymmetric constructions within type II RCFT's, see \cite{Dijkstra:2004cc,Anastasopoulos:2006da}. The first local (bottom-up) constructions were discussed in \cite{Antoniadis:2000ena,Aldazabal:2000sa}.}, the next step is to investigate finer details, such as the Yukawa couplings, which can be extracted from string amplitudes \cite{Cvetic:2003ch,Abel:2003vv,Lust:2004cx,Bertolini:2005qh} and are typically suppressed by worldsheet instantons
 \cite{Aldazabal:2000cn,Cremades:2003qj}. While worldsheet instantons can in principle account for the observed mass scales, a large amount of fine-tuning is required to obtain realistic mass hierarchies and mixings. 
 Furthermore, they are of no help in generating couplings that are perturbatively forbidden due to the violation of global $U(1)$'s, which are remnants of the generalized Green-Schwarz mechanism. In the absence of other effects, the perturbative absence of Yukawa couplings often gives rise to massless fermions, which is a phenomenological disaster.

Recently, it has been realized that D-brane instantons can break these
global symmetries and generate otherwise forbidden couplings \cite{Blumenhagen:2006xt,Haack:2006cy,Ibanez:2006da,Florea:2006si}\footnote{For a recent review on the D-instanton effects, see \cite{Blumenhagen:2009qh} and also \cite{Akerblom:2007nh,Cvetic:2007sj,Bianchi:2009ij}.}. In Type IIA compactifications, the relevant objects are so-called E2-instantons, which wrap a three-cycle in the internal manifold and are point-like in spacetime. An
instanton of this type can generate a perturbatively forbidden superpotential
term only if it compensates for the global $U(1)$ charges carried by the
forbidden coupling. This instanton induced coupling is suppressed
by the classical action of the instanton, which depends on the volume of the three-cycle that the instanton wraps. Thus, one naturally obtains a hierarchy between
perturbatively realized couplings and non-perturbatively induced couplings, and more generally between two couplings generated by instantons carrying different global $U(1)$
charges.


Often times, the same instanton which generates a perturbatively forbidden, but desired, coupling also gives rise to phenomenological drawbacks, such as the generation of R-parity violating couplings or a $\mu$-term which is too large. In \cite{Cvetic:2009yh}, the authors present the entire class
of globally consistent three-stack and four-stack MSSM D-brane quivers which give rise to the MSSM superpotential, induced perturbatively or non-perturbatively, and furthermore  
satisfy bottom-up constraints that ensure the absence of
major phenomenological drawbacks. The constraints ensure the
absence of R-parity violating couplings on both the perturbative and non-perturbative level, a 
$\mu$-term of the right order, and that all MSSM fermions are massive.
Moreover, the quivers  are required to exhibit a mechanism which explains the smallness of the neutrino masses.


In \cite{Anastasopoulos:2009mr}, the authors investigated how mass hierarchies
arise in intersecting brane models,
another very important phenomenological feature. In generating mass hierarchies,
they presented and utilized the mechanism of family splitting, where different
matter field families arise from different sectors in the D-brane spectrum. In
such a case, some entries in a given Yukawa texture might be perturbatively
allowed, whereas others are perturbatively forbidden and must be generated
via D-instantons or higher order couplings containing the VEV's of standard
model singlets. This mechanism allows for Yukawa textures of a variety of different forms, which generically allow for interesting mass hierarchical structures.


Another mechanism that naturally gives rise to mass hierarchies between different families in D-brane compactifications was presented   
in \cite{Cvetic:2009new} (see also \cite{Blumenhagen:2007zk,Ibanez:2008my,Cvetic:2009yh}). In this mechanism, non-perturbative effects generate a factorizable Yukawa texture, $Y^{IJ}\sim Y^IY^J$, which only gives mass to one family. In order to induce masses for the remaining two families, the presence of two other instantons which wrap different cycles, but have the same intersection pattern, is required. Due to the fact that the instantons wrap different cycles in the internal manifold, they can account for the observed hierarchies. 

Utilizing both family splitting and factorizable Yukawa textures for generating mass hierarchies,  the four-stack quivers of \cite{Cvetic:2009yh} were further examined in \cite{Cvetic:2009new}, with the conclusion that none exhibit proper\footnote{We refer to a proper mass
hierarchy as one that exhibits three families for the up-flavor quarks, down-flavor quarks, and charged leptons, as well as a t-quark which is much heavier than all other MSSM matter fields. Also note that, from this point on, we often refer to up-flavor and down-flavor quarks as up-quarks and down-quarks, for the sake of brevity.}
 mass hierarchies. Furthermore, in \cite{Cvetic:2009new}, five-stack models were investigated and one of a few quivers with proper hierarchies was presented. This quiver contains a dangerous dimension $5$ proton decay operator $u_R u_R d_R E_R$ which is generated by an instanton, but the associated suppression factor is high enough to ensure that the rate of proton decay due to the operator is below the current experimental bound. Furthermore, a tension was noticed in the Madrid embedding between down-flavor quark mass hierarchies obtained via factorizable Yukawa textures and the presence of proton decay operators.


In this note, we extend the analysis performed in \cite{Cvetic:2009yh} by embedding the constraints arising from proton decay via dimension $5$ operators. We present all 
four-stack quivers in the Madrid embedding which pass this further constraint and will see that none of the four stack quivers give rise to the desired Yukawa textures. We also further investigate the tension between the presence of these dimension $5$ operators and the presence of right-handed quarks transforming as antisymmetric representations of $SU(3)_C$, which can often often be utilized for giving mass hierarchies to either the up-quarks or down-quarks via factorizable Yukawa textures. It turns out that in any globally viable MSSM hypercharge embedding, the absence of right-handed quarks transforming as antisymmetric representations of $SU(3)_C$ is \emph{sufficient} to ensure that these dangerous operators are perturbatively absent and are not non-perturbatively generated by an instanton whose presence is required to generate a perturbatively forbidden, but desired, Yukawa coupling. We discuss these ideas in the context of a five-stack quiver, first presented in \cite{Cvetic:2009new}, which exhibits proper mass hierarchies and no rapid proton decay.

This note is organized as follows.  In section \ref{sec hierarchies}, we present two mechanisms for obtaining mass hierarchies in orientifold compactifications. In section \ref{sec constraints}, we explicitly discuss the top-down and bottom-up constraints that we require quivers to satisfy. In section \ref{sec proton}, we present current bounds on the suppression factors of dimension $5$ operators due to the bound on the lifetime of the proton. We also discuss the interplay between mass hierarchies and proton decay. In section \ref{sec madrid}, we present all four-stack quivers in the Madrid embedding which are consistent with the strong constraints laid out in section \ref{sec constraints} and analyze whether or not these quivers give rise to realistic hierarchies via the two mechanisms discussed in section \ref{sec hierarchies}. In section \ref{sec setup}, we present an analysis of proton decay in the aforementioned five-stack quiver with proper mass hierarchies.

\section{Mass Hierarchies in Orientifolds\label{sec hierarchies}}
A number of mechanisms exist which give rise to mass hierarchies in orientifold compactifications. We review\footnote{For a more in depth discussion of these mechanisms, see \cite{Cvetic:2009new}.} two of them here, both of which are independent of geometric specifics of the compactification manifold. A general feature exhibited by these compactifications is that they exhibit global $U(1)$'s which are remnants of the Green-Schwarz mechanism and often forbid phenomenologically desired Yukawa couplings. In such a case, these forbidden couplings may be generated by D-instantons \cite{Blumenhagen:2006xt, Ibanez:2006da, Florea:2006si} or higher order couplings containing the VEV's of standard model singlets $\phi_i$ \cite{Dutta:2005bb,Chen:2008rx,Anastasopoulos:2009mr}, provided that the global $U(1)$ charges carried by these effects precisely cancel the charges of the forbidden Yukawa coupling. The suppression factors associated with these effects are $e^{-S^{cl}_{E2}}$, where $S^{cl}_{E2}$ is the classical action of the instanton, and $\prod_i \frac{<\phi_i>}{M_S}$, respectively. They can account for the suppression of particular Yukawa couplings and thus might give an explanation for the fermion mass hierarchies of the MSSM. While both mechanisms we discuss ultimately utilize these suppression factors
to obtain mass hierarchies, they are quite different from a physical and mathematical point of view.

In \cite{Anastasopoulos:2009mr}, the authors utilized the mechanism of family splitting to generate mass hierarchies. In such a case, the fact that different families arise from different sectors in the D-brane spectrum gives their Yukawa texture entries different charges under the global $U(1)$'s. This
might cause some entries of a given Yukawa texture $Y^{IJ}$ to be perturbatively allowed, while others must be generated by
a D-instanton or via higher order couplings. For example, in the four-stack quiver with the Madrid hypercharge embedding given by
\begin{equation}
U(1)_Y=\frac{1}{6}U(1)_a+\frac{1}{2}U(1)_c+\frac{1}{2}U(1)_d\,\,,
\end{equation}
an MSSM quiver might contain two families of the right-handed down-flavor quarks, $d_R$, which transform as $(\ov{a},c)$, one family of $d_R$ which transforms as $(\ov{a},d)$, $H_d$ transforming as $(b,\ov c)$, and all three families of $q_L$ transforming as $(a,b)$. In such a case, the down quark mass matrix would take the form 
\begin{align}
M=\left(
\begin{array}{ccc}
A &  A &  B \\
A & A& B  \\
A & A & B
\end{array}
\right)
\,\,\,\,\,\,\,\,\,
\label{eq: Yukawa textures quarks}
\end{align}
where A and B carry different global $U(1)$ charge, and thus are generated by different instantons\footnote{The perturbatively missing couplings might also be generated via higher order couplings, where the SM singlets acquire a VEV. The consequences, however, are analogous to those of D-instantons. From now on we assume the perturbatively missing couplings are generated by D-instantons.}. These instantons generically carry different suppression factors, since they wrap different cycles in the internal manifold. In this case we would expect two down-quark masses to have $m\simeq A$ and one to have $m\simeq B$. Note well that this family splitting mechanism can be implemented equally well when the forbidden couplings are generated either by D-instantons or higher order couplings containing the VEV's of standard model singlets.

Unlike the previous mechanism, the mechanism presented below is purely non-perturbative. In \cite{Cvetic:2009new}, the authors utilize the fact that D-instanton effects often give rise to factorizable Yukawa textures \cite{Blumenhagen:2007zk,Ibanez:2008my,Cvetic:2009yh} to 
explain fermion mass hierarchies. We illustrate this mechanism with a concrete example. Consider three $U(1)$ branes that exhibit the intersection pattern\footnote{Our sign convention is that positive intersection number $I_{ab}=K$ correponds to K fields transforming as $(a,\ov{b})$. Elsewhere in the literature, the opposite convention is sometimes chosen.}
\begin{align*}
I_{ab}=K \qquad I_{ac}=0 \qquad I_{bc}=K\,\,,
\end{align*}
where we denote fields arising from the $ab$ sector as $\Phi^I$ and fields arising from the $bc$ sector as $\widetilde \Phi^I$.  The superpotential term
\begin{align*}
\Phi^I_{(1,-1,0)} \, \widetilde \Phi^J_{(0,1,-1)}
\end{align*}
is perturbatively forbidden, where the subscript denotes the charge under the global $U(1)_a$, $U(1)_b$ and $U(1)_c$, respectively. An instanton with the intersection pattern\footnote{Note that, for brevity's sake, we have omitted discussion of the instanton without vector-like zero modes, since it does not give rise to a factorizable Yukawa texture. For a more complete discussion of this example, see \cite{Cvetic:2009new}.}
\begin{align}
I_{E2a}=1 \qquad I_{E2b}=0 \qquad I_{E2c}=-1 \qquad I^{{\cal N}=2}_{E2b}=1 \label{eq: instanon factorizable intersection pattern}
\end{align}
carries global $U(1)$ charge $Q_{E2}(a)=-1$, $Q_{E2}(b)=0$ and $Q_{E2}(c)=1$. Such an instanton exhibits four charged zero modes, namely $ \ov \lambda_a$, $\lambda_b$, $\ov\lambda_b$ and $\lambda_c$. Its action generically takes the form
\begin{align*}
S_{E2}= S^{cl}_{E2} + Y^{IJ} \,\ov \lambda_a \,\Phi^I\,\widetilde \Phi^J \,\lambda_c + Y^{I} \,\ov \lambda_{a} \, \Phi^{I}   \lambda_b + Y^{J}\, \ov \lambda_{b} \, \widetilde \Phi^{J} \lambda_c\,\,.
\end{align*}
The contribution to the superpotential is calculated by performing the path integral over all instanton zero modes
\begin{align*}
M_s\,\,\int d^4 x \, d^2 \theta\, d \ov \lambda_a\,  d \lambda_b\, d \ov \lambda_b \, d \lambda_c \,\,e^{-S_{E2}},
\end{align*}
 where the Grassmann variables $\lambda_b$ and $\ov \lambda_b$ prevent the $Y^{IJ}$ term in the action from contributing. The resulting
 instanton induced mass matrix is given by
\begin{align}
M^{IJ}= Y^{I} \, Y^{J} \,e^{-S^{cl}_{E2}}\, M_s\,\,.
\end{align}
Since this matrix factorizes, only one linear combination of $\Phi^I\, \widetilde \Phi ^J$ becomes massive. An additional $K-1$ instantons with the intersection pattern \eqref{eq: instanon factorizable intersection pattern} are required to ensure that each family receives a mass. The associated
masses depend on the suppression factors of the instantons, which are generically of different order, since they wrap different cycles
in the internal manifold. Thus, mass hierarchies can also be explained by non-perturbative effects which give rise to a factorizable Yukawa
texture.

\section{Top-Down and Bottom-Up Constraints\label{sec constraints}}
In this section, we briefly summarize the constraints we require D-brane quivers to satisfy. For a more detailed description, we refer the reader to \cite{Cvetic:2009yh}. We distinguish between the two different classes of constraints, \emph{top-down} and \emph{bottom-up} constraints. The former include constraints on the chiral matter field transformation behavior arising from tadpole cancellation and from the presence of a massless hypercharge $U(1)_Y$.
The latter are due to experimental observations.

\begin{itemize}
\item[$\bullet$] All of the MSSM matter fields and the right-handed neutrino, apart from the Higgs
fields, appear as chiral fields at intersections between two stacks of D-branes. Furthermore, the spectrum must contain no chiral exotics.
\item[$\bullet$] As discussed in \cite{Cvetic:2009yh}, tadpole cancellation, which is a condition on the cycles that the D-branes wrap, imposes constraints on the transformation behavior of the chiral matter. For a stack of $N_a$ D-branes  with $N_a>1$, the constraints read
\begin{align}
\#(a) - \#({\ov a})  + (N_a-4)\#( \, \Yasymm_a) + (N_a+4) \#
(\Ysymm_a)=0 \,\,,\label{eq constraint1}
\end{align}
while for $N_a=1$ it is slightly modified and takes the form
\begin{align}
\#(a) - \#({\ov a}) + 5 \# (\Ysymm_a)=0  \qquad \text{mod} \,3\,\,.
\label{eq constraint2}
\end{align}
We require that all D-brane stacks satisfy these conditions.
\item[$\bullet$] The presence of a massless $U(1)_Y$ is also a condition on the cycles that the D-branes wrap, which puts constraints on the transformation behavior of the chiral matter, given by
\begin{align}
 \sum_{x \neq a} q_x\,
N_x \#(a,{\ov x}) - \sum_{x \neq a} q_x\, N_x \#(a,x) = q_a\,N_a \,\Big(\#(\Ysymm_a) + \# (\, \Yasymm_a)\Big) \,\,
\label{eq massless constraint non-abelian}
\end{align}

for $N_a>1$ and
\begin{align}\sum_{x \neq a} q_x\, N_x \#(a,{\ov x}) -
\sum_{x \neq a} q_x\, N_x \#(a,x)= q_a\,\frac{\#(a) - \#({\ov a}) + 8 \#
(\Ysymm_a)}{3}  
\label{eq massless constraint abelian}
\end{align}
for $N_a=1$.
We require that all D-brane stacks satisfy these conditions.
\item[$\bullet$]We require that each quiver exhibits non-zero masses for all three families of the up-quarks, down-quarks, and charged leptons. 
\item[$\bullet$] We forbid R-parity violating couplings on both the perturbative and non-perturbative level.
\item[$\bullet$] We require that there is no instanton needed to generate a Yukawa coupling which also generates a tadpole $N_R$.
\item[$\bullet$]
We rule out setups which lead to a large family mixing in the quark Yukawa couplings \cite{Ibanez:2008my,Leontaris:2009ci,Cvetic:2009yh}.
\item[$\bullet$]  The D-brane quiver must allow for a mechanism which gives a $\mu$-term of the observed order.
\item[$\bullet$] We require that the D-brane quiver exhibits a mechanism which accounts for the smallness of the neutrino masses
\cite{Blumenhagen:2006xt,Ibanez:2006da,Cvetic:2007ku,Ibanez:2007rs,Antusch:2007jd,Cvetic:2007qj,Cvetic:2008hi}
.
\item[$\bullet$] We require that the $t$-quark mass is at least two orders of magnitude larger than the masses of any other MSSM matter field.  
\item[$\bullet$] If present, we require that the suppression of the dimension $5$ operators
$q_L q_L q_L L$ and $u_R u_R d_R E_R$ is sufficient to satisfy the current bounds on proton decay, discussed in section \ref{sec proton}.
\end{itemize}

Fore more details on the constraints, we refer  the reader to \cite{Cvetic:2009yh,Cvetic:2009new}. In constrast to \cite{Cvetic:2009yh}, here we explicitly impose the proton decay constraint. Also, rather than marking them with a $\clubsuit$, we remove all quivers with too much family mixing. 

\section{A Discussion of Proton Decay\label{sec proton}}
In this section, we discuss further the constraints which ensure the absence of rapid proton decay due to dimension $5$ operators. Such disastrous effects arise when the dimension $5$ operators 
\begin{align}
\frac{\kappa}{M_s}  \,\,q_L\, q_L \,q_L \,L \qquad \text{and} \qquad \frac{\kappa'}{M_s}\,\, u_R \, u_R \, d_R \,E_R\,\,
 \end{align}
 are present in the superpotential with inadequate suppression. Choosing $M_s\simeq10^{18}$  $GeV$, the experimental upper bound on the proton lifetime requires that $\kappa$ and $\kappa'$ satisfy 
 \begin{align}
 \kappa, \kappa' \leq 10^{-8} \,\,,
 \label{eq bounds on kappa}
\end{align} 
 as given, for example, in \cite{Hinchliffe:1992ad}. Thus, any quiver in which either of these couplings is perturbatively realized exhibits rapid proton decay, since $\kappa$ or $\kappa'$  is O(1) in the absence of extreme fine-tuning due to worldsheet instanton suppression. Such a quiver is ruled out as unrealistic.
 
 Fortunately, the operator $q_L q_L q_L L$ can never be perturbatively realized, since it will always be charged under the global symmetry $U(1)_a$ arising from the color D-brane stack. The coupling $u_R u_R d_R E_R$, on the other hand, has a chance of being uncharged under this symmetry, since in some hypercharge embeddings one of the right-handed quarks might transform as antisymmetric of $SU(3)_C$. This frequently occurs in quivers with a Madrid-type hypercharge embedding, where the $d_R$ can transform in this fashion if it arises at the intersection of the $a$-brane with its orientifold image.
 
 Furthermore, even if these couplings are perturbatively forbidden, they might be induced by an instanton which is required to generate one of the perturbatively forbidden, but desired, Yukawa couplings. In each case, a careful analysis of the suppression factors associated with such instantons is required to determine whether or not the bounds on $\kappa$ and $\kappa'$ in equation (\ref{eq bounds on kappa}) are satisfied.
 A general rule of thumb, though, is that the quiver does not exhibit rapid proton decay if the instanton which induces one of the dimension $5$ operators is required to induce
 the $\mu$-term or a  Dirac neutrino mass with Dirac-like suppression\footnote{A ``Dirac-like suppression" is one that explains the smallness of the neutrino masses without
 employing the seesaw mechanism \cite{Cvetic:2008hi} and is expected to be of the order $10^{-14}-10^{-11}$.  }, since the suppression associated with these instantons is very high. On the other hand, instantons required to induce Yukawa couplings for the up-quarks, down-quarks, and charged leptons usually are not suppressed enough to satisfy the bounds.
Note that in the absence of right-handed quarks transforming as antisymmetrics, the $U(1)_a$ charge of both dimension $5$ operators is different than that of any instanton required to generate an MSSM Yukawa coupling, and thus one does not have to worry about a desired Yukawa coupling inducing instanton generating the dangerous dimension 5 proton decay operators.
 
We emphasize that these statements apply to all globally viable\footnote{Many MSSM hypercharge embeddings exist, though only a small subset are able to satisfy the constraints due to tadpole cancellation and masslessness of $U(1)_Y$. The latter are what we mean by ``globally viable" hypercharge embeddings, which are dicussed in an appendix in \cite{Cvetic:2009new}.} hypercharge embeddings, rather than just the Madrid embedding, which was the only case discussed in \cite{Cvetic:2009new}. Very precisely, this means that in any globally viable MSSM hypercharge embedding, the absence of right-handed quarks transforming as antisymmetrics of $SU(3)_C$ is \emph{sufficient} to ensure that these dangerous operators are perturbatively absent and are not non-perturbatively generated by an instanton whose presence is required to generate a perturbatively forbidden, but desired, Yukawa coupling. 
For the purpose of obtaining mass hierarchies, however, right-handed quarks transforming as antisymmetrics are quite useful, since it is only in this case that the corresponding Yukawa texture might be factorizable. Thus, there is a tension between a quiver obtaining quark mass hierarchies via a factorizable Yukawa texture and it not exhibiting dangerous dimension 5 operators which lead to rapid proton decay.

\section{Semi-Realistic Four-Stack Madrid Quivers \label{sec madrid}}
In \cite{Cvetic:2009yh}, the authors presented all four-stack D-brane quivers which realize the MSSM and satisfy most\footnote{Recall that, compared to \cite{Cvetic:2009yh}, here we also impose the constraints which ensure the absence of dangerous dimension $5$ operators. Therefore, the number
of quivers listed above which satisfy the constraints in \cite{Cvetic:2009yh} would be further cut down by this additional constraint.} of the constraints discussed in section \ref{sec constraints}.  Of the roughly
$10,000$ setups that satisfied the constraints due to tadpole cancellation and the presence of a massless $U(1)_Y$, only about $70$ pass the phenomenological bottom-up
constraints. The most fruitful hypercharge embedding is the Madrid embedding\footnote{Note that, in constrast to \cite{Cvetic:2009yh}, we have chosen the Madrid embedding to have all plus signs. This is for consistency with the extended Madrid embedding convention in \cite{Cvetic:2009new} and here. The sign in question is on the coefficient of $U(1)_d$, and quivers can be mapped from one convention
to the other simply by exchanging $d \leftrightarrow d'$ as stacks and sending $d \leftrightarrow \ov{d}$ in the transformation behavior.},
\begin{equation}
U(1)_Y= \frac{1}{6} U(1)_a +\frac{1}{2} U(1)_c +\frac{1}{2}U(1)_d,
\label{eqn madrid}
\end{equation}
which accounts for $45$ of those solutions. This success is not entirely surprising, given that, in this embedding, a given MSSM matter field might transform in a number of different ways.
For this hypercharge, the potential 
transformation behavior of the MSSM matter fields is given by

\begin{align*}
q_L\,:&\qquad (a,\ov{b}),\,\,\,\,(a,b)\\
u_R\,:&\qquad (\ov{a},\ov{c}),\,\,\,\,(\ov{a},\ov{d})\\
d_R\,:&\qquad {\Yasymm}_a,\,\,\,\,(\ov{a},c),\,\,\,\,(\ov{a},d)\\
L\,\,:&\qquad (b,\ov{c}),\,\,\,\,(\ov{b},\ov{c}),\,\,\,\,(\ov b,\ov{d}),\,\,\,\,(b,\ov{d})\\
E_R\,:&\qquad (c,\ov{d}),\,\,\,\,{\Ysymm}_c,\,\,\,\,{\Ysymm}_d\\
N_R\,:&\qquad {\Yasymm}_b,\,\,\,\,\ov{\Yasymm}_b\,,\,\,\,(c,\ov d),\,\,\,\, (\ov c, d)  \\
H_u\,:&\qquad (\ov b,c),\,\,\,\,(b,c),\,\,\,\,(b,d),\,\,\,\,(\ov{b},d)\\
H_d\,:&\qquad (b,\ov{c}),\,\,\,\,(\ov{b},\ov{c}),\,\,\,\,(\ov b,\ov d),\,\,\,\,(b,\ov d)\,\,.
\end{align*}


We extend the analysis of \cite{Cvetic:2009yh} by requiring that D-brane quivers in the Madrid embedding do not give rise to rapid proton decay. Furthermore, in contrast to \cite{Cvetic:2009yh}, we explicitly remove quivers which exhibit an unrealistic CKM matrix. It is often quite obvious from the Yukawa textures $Y_{q_LH_uU_R}$ and $Y_{q_LH_dd_R}$ whether or not this is the case. For example, the quark textures might take the form
\begin{align}
Y_{q_LH_uu_R}=\left(
\begin{array}{ccc}
P &  P &  P \\
A & A & A  \\
A & A & A
\end{array}
\right)
\,\,\,\,\,\,\,\,\,
Y_{q_LH_dd_R}=\left(
\begin{array}{ccc}
B &  B &  B \\
P & P & P  \\
P & P & P
\end{array}
\right),
\label{eq: Yukawa textures quarks}
\end{align}
where the entries P correspond to perturbatively realized couplings and the entries A and B correspond to perturbatively forbidden couplings that are generated with suppression by instantons. Since the perturbatively realized couplings are generically of higher order, matrix structures of this form immediately imply that there is too much family mixing, and thus an unrealistic CKM matrix.

With regard to proton decay, it turns out that any four-stack quiver in the Madrid embedding satisfying the previously discussed top-down and bottom-up constraints exhibits $u_Ru_Rd_RE_R$ at the perturbative level if and only if it has a right-handed down-flavor quark, $d_R$, transforming as antisymmetric of $SU(3)_C$. For all other quivers none of the dangerous dimension 5 operators is realized perturbatively. Moreover, in these quivers no instanton whose presence is required to induce some of the perturbatively missing MSSM couplings generates the dimension 5 operators $q_L q_L q_L L$ or $u_Ru_Rd_RE_R$. In Table \ref{table}, we display all setups which satisfy all the constraints presented in section \ref{sec constraints}. We emphasize again that there is no quiver with $d_R$ transforming as an antisymmetric of $SU(3)_c$, realized here as $\Yasymm_a$, due to the interplay between antisymmetrics and rapid proton decay.

Note that, in comparison to \cite{Cvetic:2009yh}, which did not omit quivers with too much family mixing or rapid proton decay, there are far fewer solutions here. One drastic consequence of the additional constraints is that the mass hierarchy of the surviving quivers tends to be worse. Specifically, one quiver of \cite{Cvetic:2009yh}, discussed explicitly in \cite{Cvetic:2009new}, nearly exhibited proper mass hierarchy, with the only deficiency being the existence of two up-quark hierarchies, rather than three. This quiver does not survive the additional constraints, and moreover all of the quivers in Table \ref{table} have additional mass hierarchical deficiencies. These deficiencies might include a perturbatively realized down-quark or charged lepton coupling, or only two hierarchies for the down-quarks or charged leptons. This further motivates the examination of five-stack quivers.

\begin{table}[h]
\hspace{-.75cm}
\scalebox{.55}{
\begin{tabular}{|c|c|c|c|c|c|c|c|c|c|c
		|c|c|c|c|c|c|c|c|c|c|c|}\hline
		\multirow{2}{*}{Solution}&\multicolumn{2}{|c}{$q_L$} & \multicolumn{2}{|c}{$d_R$} & \multicolumn{2}{|c}{$u_R$} & \multicolumn{3}{|c}{$L$} & \multicolumn{3}{|c}{$E_R$}
		& \multicolumn{4}{|c}{$N_R$} & \multicolumn{4}{|c}{$H_u$} & \multicolumn{1}{|c|}{$H_d$}\\ \cline{2-22}
		&$(a,b)$ & $(a,\ov{b})$ & $(\ov{a},c)$ & $(\ov{a},d)$ & $(\ov{a},\ov{c})$
		& $(\ov{a},\ov d)$ & $(b,\ov{c})$  & $(b, \ov d)$ & $(\ov{b}, \ov d)$ & $(c,d)$
		& ${\Ysymm}_c $ & $\Ysymm_d $ & $\Yasymm_b $ & $\ov{\Yasymm}_b $ & $(c,\ov d) $ & $(\ov{c},d) $
		& $(b,c)$ & $(\ov{b},c)$ & $(b,d)$ & $(\ov{b},d) $ & $(\ov{b},\ov{c})$ \\
		\hline\hline			
1$^{\dagger}$&3&0&3&0&0&3&0&0&3&0&0&3&2&0&0&1&0&0&0&1&1\\ \hline
2&3&0&3&0&0&3&0&0&3&1&0&2&2&0&1&0&0&0&0&1&1\\ \hline
3&3&0&3&0&2&1&0&0&3&2&1&0&2&0&1&0&0&0&0&1&1\\ \hline
4&3&0&3&0&2&1&0&0&3&0&2&1&2&0&1&0&0&0&0&1&1\\ \hline
5&3&0&3&0&3&0&0&0&3&2&1&0&2&0&0&1&0&1&0&0&1\\ \hline
6&3&0&3&0&3&0&0&0&3&0&2&1&2&0&0&1&0&1&0&0&1\\ \hline
7&3&0&3&0&3&0&0&0&3&1&2&0&2&0&1&0&0&1&0&0&1\\ \hline
8$^\heartsuit$&0&3&0&3&0&3&3&0&0&2&0&1&0&3&0&0&0&0&1&0&1\\ \hline
9$^\heartsuit$&0&3&0&3&0&3&3&0&0&0&1&2&0&3&0&0&0&0&1&0&1\\ \hline
10&0&3&0&3&1&2&3&0&0&3&0&0&0&3&0&0&1&0&0&0&1\\ \hline
11&0&3&0&3&1&2&3&0&0&1&1&1&0&3&0&0&1&0&0&0&1\\ \hline
12$^{\dagger}$&0&3&0&3&3&0&3&0&0&0&3&0&0&3&0&0&1&0&0&0&1\\ \hline
\end{tabular}}
\caption{Spectra for the setups with $U(1)_Y= \frac{1}{6} U(1)_a +\frac{1}{2} U(1)_c +\frac{1}{2}U(1)_d$.
\label{table}
}
\end{table}

\section{A Quiver with Proper Mass Hierarchies and No Dangerous Proton Decay\label{sec setup}}
Though the quivers in the previous section are compatible with all bottom-up constraints arising from experimental observations, they all fail to exhibit proper mass hierarchies. Extending the quiver by an additional $U(1)$ D-brane stack might give rise to setups which exhibit proper hierarchies. This possibility was examined in \cite{Cvetic:2009new}, where it was shown that there are only three possible five-stack hypercharge embeddings which might possibly give rise to the experimentally observed inter- and intra-family mass hierarchies while satisfying all the top-down and bottom-up constraints. The most promising of these is the extended Madrid embedding,

\begin{equation}
U(1)_Y = \frac{1}{6} U(1)_a + \frac{1}{2} U(1)_c + \frac{1}{2} U(1)_d + \frac{1}{2} U(1)_e\,\,.
\end{equation}

With this embedding, there are a few quivers which not only exhibit proper mass hierarchies, but also overcomes the serious issue of the dangerous dimension $5$ operators which lead to rapid proton decay.  We now present one such quiver\footnote{A detailed discussion in \cite{Cvetic:2009new} showed that this quiver exhibits proper mass hierarchy. Here, instead of discussing mass hierarchical specifics, we focus on the interplay between mass hierarchies and proton decay.}, first discussed in \cite{Cvetic:2009new}, where the origin and transformation of the MSSM matter fields is given in Table \ref{table spectrum for 5-stack quiver setup 3}. 

\begin{table}[h] \centering
\begin{tabular}{|c|c|c|c|c|}
\hline
 Sector & Matter Fields &  Transformation & Multiplicity & Hypercharge\\
\hline \hline
  $ab$                            & $q_L$  & $(a,\overline{b})$ & $1$& $\frac{1}{6}$ \\
 \hline
 $ab'$                            & $q_L$  & $(a,b)$ & $2$& $\frac{1}{6}$ \\
\hline
 $ac'$                            & $u_R$  & $(\overline{a},\overline{c})$  & $2 $ & $-\frac{2}{3}$ \\
\hline
 $ad'$                            & $u_R$  & $(\overline{a}, \ov d)$  & $1 $ & $-\frac{2}{3}$ \\
\hline
$aa'$ & $d_R$ & ${\Yasymm}_a$ & $3$& $\frac{1}{3}$  \\
\hline
$bc'$                            & $H_u$  & $(b,c)$ & $1$ & $\frac{1}{2}  $  \\
\hline
$bd'$                            & $L$  & $(\ov b,\overline{d})$  & $3$& $-\frac{1}{2}$ \\
\hline
$be'$                            & $H_d$  & $(\overline{b},\overline{e})$ & $1$ & $\frac{1}{2}  $  \\
\hline
$ce'$                            & $E_R$  & $(c,e)$  & $2 $ & $1$ \\
\hline
$ce$                            & $N_R$  & $(\ov c, e)$  & $1 $ & $0$ \\
\hline
$dd'$                            & $E_R$  & $\Ysymm_d$  & $1 $ & $1$ \\
\hline
$de$                            & $N_R$  & $(  \ov d, e)$  & $2 $ & $0$ \\
\hline
\end{tabular}
\caption{\small {A quiver in the extended Madrid embedding.} }
\label{table spectrum for 5-stack quiver setup 3}
\end{table}

As discussed in section \ref{sec proton}, the absence of right-handed quarks transforming as antisymmetric representations of $SU(3)_C$ is \emph{sufficient} to ensure that the dangerous dimension $5$ operators $q_Lq_Lq_LL$ and $u_R u_R d_R E_R$ are perturbatively absent and are not  generated  by an instanton whose presence is required to induce one of the perturbatively missing, but desired, Yukawa couplings. Thus, as is evident from the spectrum, this quiver might potentially exhibit rapid proton decay. In fact, the presence of this transformation behavior for the right-handed down-flavor quarks is precisely the reason why they have three mass hierarchies in this quiver. A closer look reveals that both operators, $q_L q_L q_L L$ and $u_R u_R d_R E_R$, are absent on perturbative level. Moreover, the coupling $q_L q_L q_L L$ is not induced by any of the instantons which are required to generate the perturbatively missing MSSM couplings. 
On the other hand, the dimension 5 operator $u_Ru_Rd_RE_R$ is generated by an instanton with the intersection pattern\footnote{Here we assume that the instanton wraps an orientifold invariant cycle and thus exhibits the right uncharged zero mode structure to give contributions to the superpotential \cite{Argurio:2007qk,Argurio:2007vq,Bianchi:2007wy,Ibanez:2007rs}.}
\begin{align*}
 I_{E2a}=0 \qquad  I_{E2b}=0  \qquad I_{E2c}=0 \qquad  I_{E2d}=-1 \qquad I_{E2e}=1\qquad I^{{\cal N}=2}_{E2c}=1\,\,.
\end{align*}
whose presence is required to induce a Dirac neutrino mass term. Whether or not this operator gives rise to rapid proton decay depends entirely on whether or not the suppression factor $\kappa'\simeq e^{-S^{cl}_{E2}}$ satisfies the bound in equation (\ref{eq bounds on kappa}). The value of the suppression factor itself is determind by which Yukawa coupling the instanton is required to induce. Here the instanton induces the Dirac neutrino mass term and thus the suppression factor is expected to be in the range $10^{-14}-10^{-11}$, which is more than enough to evade the bound on proton lifetime.  Moreover, the presence of the instanton is not required, since experiments have not yet ruled out the possibility of a massless neutrino family. In that case, the dimension $5$ operator  $u_R \, u_R\, d_R \,E_R$ would not be induced by any of the instantons generating the perturbatively missing MSSM couplings. We conclude that this quiver does not suffer from rapid proton decay, and thus provides a viable setup which gives rise to realistic phenomenology.

\vspace{1.5cm}
{\bf Acknowledgments}\\
We thank M. Ambroso, M. Bianchi, T. Brelidze, B. G. Chen, I. Garc\'ia-Etxebarria, L. E. Ib{\'a}{\~n}ez, E. Kiritsis, A. Lionetto, J. F. Morales, B. Schellekens and T. Weigand for
useful discussions. The work is supported by the DOE  Grant DOE-EY-76-02-3071, the NSF RTG grant DMS-0636606 and the Fay R. and Eugene L. Langberg
Chair.

\bibliography{rev}
\bibliographystyle{utphys}

\end{document}